\documentclass[aps,prd,twocolumn,nofootinbib,superscriptaddress,preprintnumbers]{revtex4-1}
\usepackage{amsfonts,amsmath,amssymb,bm}
\usepackage{booktabs,array}
\usepackage[utf8]{inputenc}
\usepackage{float,graphicx}
\usepackage{appendix}
\usepackage[dvipsnames]{xcolor}
\usepackage{mathtools}
\usepackage[caption=false]{subfig}

\AtBeginDocument{
\heavyrulewidth=.08em
\lightrulewidth=.05em
\cmidrulewidth=.03em
\belowrulesep=.65ex
\belowbottomsep=0pt
\aboverulesep=.4ex
\abovetopsep=0pt
\cmidrulesep=\doublerulesep
\cmidrulekern=.5em
\defaultaddspace=.5em
}

\usepackage{hyperref}
\hypersetup{
    pdftitle={Reconstructing QCD Spectral Functions with Gaussian Processes},
    colorlinks=true,     
    linkcolor=blue,      
    citecolor=blue,      
    filecolor=blue,      
    urlcolor=blue        
}
\usepackage[nameinlink]{cleveref}

\begin{document}

\title{Reconstructing QCD Spectral Functions with Gaussian Processes}

\newcommand{\HeidelbergAffiliation}{\affiliation{Institut f\"ur Theoretische Physik, Universit\"at Heidelberg, Philosophenweg 16, D-69120 Heidelberg, Germany}}
\newcommand{\EMMIAffiliation}{\affiliation{ExtreMe Matter Institute EMMI, GSI, Planckstr.~1, D-64291 Darmstadt, Germany}}
\newcommand{\HuelvaAffiliation}{\affiliation{Department of Integrated Sciences and Center for Advanced Studies in Physics,
Mathematics and Computation, University of Huelva, E-21071 Huelva, Spain}}
\newcommand{\CPTAffiliation}{\affiliation{Aix Marseille Univ, Universit\'e de Toulon, CNRS, CPT, Marseille, France}}

\author{Jan~Horak} \HeidelbergAffiliation
\author{Jan~M.~Pawlowski} \HeidelbergAffiliation \EMMIAffiliation
\author{José~Rodríguez-Quintero} \HuelvaAffiliation
\author{Jonas~Turnwald} \HeidelbergAffiliation
\author{Julian~M.~Urban} \email{urban@thphys.uni-heidelberg.de} \thanks{corresponding author.} \HeidelbergAffiliation
\author{Nicolas~Wink} \HeidelbergAffiliation
\author{Savvas~Zafeiropoulos} \CPTAffiliation

\begin{abstract}
	We reconstruct ghost and gluon spectral functions in 2+1 flavor QCD with Gaussian process regression. This framework allows us to largely suppress spurious oscillations and other common reconstruction artifacts by specifying generic magnitude and length scale parameters in the kernel function. The Euclidean propagator data are taken from lattice simulations with domain wall fermions at the physical point. For the infrared and ultraviolet extensions of the lattice propagators as well as the low-frequency asymptotics of the ghost spectral function, we utilize results from functional computations in Yang-Mills theory and QCD. This further reduces the systematic error significantly. Our numerical results are compared against a direct real-time functional computation of the ghost and an earlier reconstruction of the gluon in Yang-Mills theory. The systematic approach presented in this work offers a promising route towards unveiling real-time properties of QCD.
\end{abstract}

\maketitle


\paragraph*{Introduction.}
The resolution of many open questions in quantum chromodynamics (QCD) requires the knowledge of time-like observables and hence the computation of real-time correlation functions. Applications range from the hadronic resonance spectrum over scattering processes to transport and non-equilibrium phenomena in heavy-ion collisions. For example, the computation of the glueball spectrum via Bethe-Salpeter equations relies on the time-like propagators for gluon and ghost, both of which are reconstructed in the present work. Likewise, QCD transport coefficients used in hydrodynamic simulations can be computed diagrammatically from the real-time gluon propagator. Similarly, phenomenological QCD transport models with their underlying assumption of a quasi-particle nature of the gluon can hugely benefit in multiple ways from the present results. First of all, a reliable computation of the gluon spectral function may offer much-needed support for the quasi-particle assumption of these models, as well as give access to its limitations. Secondly, the QCD gluon spectral function itself can feature as a direct input and pivotal building block in these models. Together with further time-like correlation functions, this offers a path for a systematic quantitative improvement of phenomenological transport approaches towards first-principle transport in QCD. 

By now, Euclidean correlation functions in QCD are accessible within first-principle approaches such as lattice simulations or functional equations. In contradistinction, accessing real-time properties remains a notoriously hard task. Minkowski correlation functions may be obtained from Euclidean data via spectral reconstruction, exploiting the K\"all\'en-Lehmann (KL) representation~\cite{Kaellen1952, Lehmann1954}. This requires computing the spectral function via an inverse integral transform. In the present work, we approach the problem with Gaussian process regression (GPR). The applicability of GPR to inverse problems of this type has been discussed in~\cite{10.1093/gji/ggz520}. Specifically, it was shown how GPs can be used to obtain probabilistic models of functions for which only weighted averages are available.

We apply GPR to the reconstruction of ghost and gluon spectral functions based on recent results from 2+1 flavor lattice QCD with domain wall fermions at a pion mass of 139\, MeV~\cite{Zafeiropoulos:2019flq, Cui:2019dwv}. Furthermore, we improve the systematic error control by incorporating additional data in the infrared (IR) and ultraviolet (UV) regimes from functional renormalization group (fRG) and Dyson-Schwinger (DSE) computations in Yang-Mills theory and QCD~\cite{Cyrol:2016tym, Cyrol:2017ewj, Cyrol:2018xeq, Fu:2019hdw, Gao:2020qsj, Gao:2021wun, Horak:2021pfr}, mostly obtained within the fQCD collaboration~\cite{fQCD}.\\[-1ex]


\paragraph*{Spectral representation.}
The KL spectral representation of the two-point correlation function in momentum space reads 
\begin{equation}\label{eq:KL_rep}
    G(p_0) = \int_0^\infty \frac{\textrm{d}\omega}{\pi} \frac{\omega\, \rho(\omega)}{\omega^2 + p_0^2} = \int_0^\infty \mathrm{d}\omega\, K(p_0,\omega)\, \rho(\omega) \ ,
\end{equation}
with the KL kernel $K(p_0, \omega)$ and $\rho(-\omega) = -\rho(\omega)$. In the vacuum, the spatial momentum dependence of the propagator can be obtained via a Lorentz boost, simply by $p_0^2\to p^2$ with $p^2=p_0^2+\vec p^2$.

With \Cref{eq:KL_rep}, the spectral function is obtained from the retarded propagator via
\begin{equation}\label{eq:rho-def}
	\rho(\omega) = 2\, \mathrm{Im}\,G(-\mathrm{i}(\omega + \mathrm{i}0^+))\ .
\end{equation}
For asymptotic states, the spectral function is the probability density for (multi-)particle excitations created from the vacuum in the presence of the corresponding quantum field. Consequently, in this case the spectral function is positive semi-definite. For propagators of `unphysical' fields, such as gauge fields, the spectral representation may still hold. However, the spectral function can then also have negative parts, and the existence of a spectral representation simply constrains the allowed complex structure of correlation functions; see e.g.~\cite{Lowdon:2017gpp, Lowdon:2018uzf, Cyrol:2018xeq, Bonanno:2021squ, Horak:2021pfr}.

In this work, we reconstruct ghost and gluon spectral functions of 2+1 flavor QCD under the assumption that both admit a KL representation. It can be shown that the total spectral weight vanishes,
\begin{equation}\label{eq:gen-ozs}
	\int_0^\infty \frac{ \textrm{d}\omega}{\pi} \, \omega \rho_{A/c}(\omega) = 0 \ ,
\end{equation}
respectively for both the ghost and gluon spectral functions, $\rho_c$ and $\rho_A$. For the gluon, this is the well-known Oehme-Zimmermann superconvergence (OZS) condition~\cite{Oehme:1979bj, Oehme:1990kd}; for recent discussions with general fields, see~\cite{Cyrol:2018xeq, Bonanno:2021squ, Horak:2021pfr}. These works also include a treatment of the analytic low-frequency behavior of continuous parts of the spectral functions, initiated in~\cite{Cyrol:2018xeq}.

A general spectral function $\rho$ consists of a continuous part $\tilde\rho$ and a sum of particle and resonance peaks (proportional to the $\delta$-function and its derivatives). In this work, we assume that the gluon spectral function only consists of a continuous part $\rho_A=\tilde\rho_A$ satisfying \Cref{eq:gen-ozs}. This is the generic structure suggested by all functional equations describing the gluon propagator due to the ghost being massless. While derivatives of $\delta$-functions are formally also allowed, we exclude these structures from our ansatz due to the absence of a generic mechanism generating the required roots of the inverse gluon propagator on the real momentum axis. In turn, due to the $1/p^2$ behavior of the Euclidean lattice ghost propagator in the IR, the associated spectral function exhibits a particle peak at vanishing frequency in addition to its continuous part, i.e.
\begin{equation} \label{eq:rho-ghost}
    \rho_c(\omega) = \frac{\pi}{Z_c} \frac{\delta(\omega)}{\omega} +\tilde \rho_c(\omega)\ ,\ 
    \int_0^\infty \!\frac{\textrm{d}\omega}{\pi} \, \omega \,\tilde \rho_{c}(\omega) = -\frac{1}{Z_c}\ ,
\end{equation}
where $\delta(\omega)/\omega$ has to be understood as a limiting process $\delta(\omega - m) / \omega$ with $m \to 0^+$. Evidently, for $Z_c = 1$ and ${\tilde\rho_c = 0}$ the ghost propagator reduces to the classical one.

Euclidean correlators obtained from lattice simulations are generally only available in terms of discrete sets of observations $G_i$ at $N_G$ Euclidean momenta $p_i$ with finite precision. Relating the results to the associated Minkowski propagators via \Cref{eq:rho-def} is problematic; see e.g.~\cite{Cuniberti:2001hm, Burnier:2011jq}. In such a numerical setup the analytic continuation via $p \to -\mathrm{i}(\omega + \mathrm{i} 0^+)$ is ill-conditioned, since further assumptions about the complex structure need to be made. Instead, the usual strategy is the numerical inversion of the integral transformation. A variety of approaches has been explored to tackle this issue, such as the maximum entropy method~\cite{Jarrell:1996rrw, Asakawa:2000tr, Haas:2013hpa}, Bayesian inference techniques~\cite{Burnier:2013nla, Rothkopf:2016luz}, suitable expansions in functional spaces~\cite{Cuniberti:2001hm, Burnier:2011jq, Cyrol:2018xeq, Fei_2021, fei2021analytical}, Padé-type approximants~\cite{Binosi:2019ecz, Falcao:2020vyr}, Tikhonov regularization~\cite{Ulybyshev:2017ped, Dudal:2019gvn, Dudal:2021gif}, neural networks~\cite{Fournier_2020, Yoon_2018, Kades:2019wtd, Zhou:2021bvw}, and kernel ridge regression~\cite{arsenault2016projected, Offler:2019eij}. Alternative approaches based on the existence of complex conjugate poles have also been considered, see e.g.~\cite{Capri:2015mna, Siringo:2016vrv, Hayashi:2018giz, Binosi:2019ecz, Dudal:2019aew, Kondo:2019ywt, Hayashi:2020few, Hayashi:2021nnj, Hayashi:2021jju}, but are orthogonal to the present work.\\[-1ex]


\paragraph*{Reconstruction with GPR.}
Starting from early developments in the context of geostatistics in the 1950s~\cite{krige1951statistical}, today GPR is widely employed in a variety of settings for the probabilistic modeling of functions from a finite number of observations; see~\cite{kanagawa2018gaussian, liu2019gaussian} for reviews and~\cite{10.5555/1162254} for a modern textbook account. Recently, the method has been applied to the reconstruction of parton distribution functions from lattice QCD~\cite{Alexandrou:2020tqq}. In this section, we summarize the main ingredients for spectral reconstruction with GPR based on the developments reported in~\cite{10.1093/gji/ggz520}. A short introduction to GPR for function prediction as well as further details and references are provided in \Cref{app:gpr}. 

We assume our knowledge of the spectral function $\rho(\omega)$ to be described by a GP, written as
\begin{equation}\label{eq:rho-gp}
	\rho(\omega) \sim \mathcal{GP}(\mu(\omega), C(\omega, \omega')) \ ,
\end{equation}
where $\mu(\omega),C(\omega,\omega')$ denote the mean and covariance functions. Importantly, in this approach we do not restrict the space of possible solutions by choosing a specific functional basis, which often leads to spurious artifacts in the reconstruction in order to compensate for unrepresentable features. Instead, the GP defines a distribution over families of functions with rather generic properties, specified via the kernel parametrization described below.

The KL integral in \Cref{eq:KL_rep} is a linear transformation that preserves Gaussian statistics. Hence, given \Cref{eq:rho-gp} one may obtain statistical predictions $G_i$ at $N_G$ specified momenta $p_i$ as
\begin{equation}\label{eq:prop-dist}
\begin{aligned}
	G_i &\sim \mathcal{N} \left(\int \textrm{d}\omega\ K(p_i, \omega) \mu(\omega), \right. \\
	& \left. \int \textrm{d}\omega\, \textrm{d}\omega' K(p_i, \omega) C(\omega, \omega') K(p_j, \omega') \right) \\ 
	&\equiv \mathcal{N}\big(\tilde\mu_i,\tilde C_{ij}\big)\ .
\end{aligned}
\end{equation}
Here, $\mathcal{N}$ denotes a multivariate normal distribution, to be distinguished from distributions over function space denoted by $\mathcal{GP}$. Statistical uncertainties associated with individual prediction points $\tilde{\mu}_i$ may be computed from the diagonal of the covariance matrix as $\tilde{\sigma}_i = \sqrt{\tilde{C}_{ii}}$.

\begin{figure*}
	\centering
	\subfloat[]{%
		\includegraphics{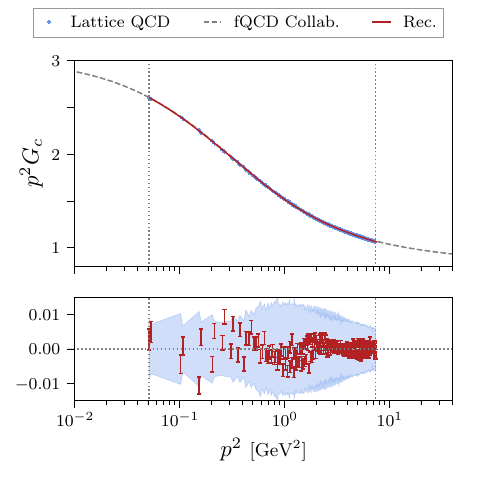}\label{fig:ghost-dressing}}
	\hfill
	\subfloat[]{%
		\includegraphics{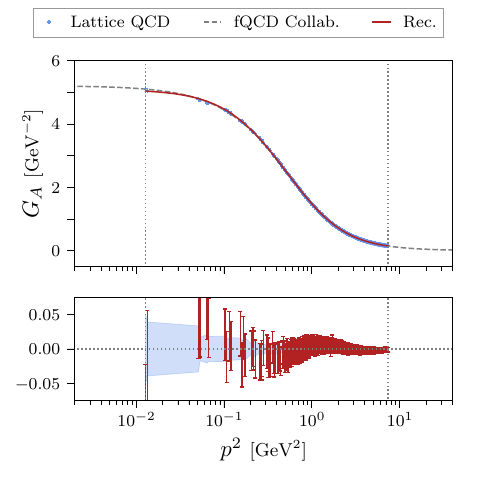}\label{fig:gluon-propagator}}%
	\caption{Plots showing the ghost dressing function (a) and gluon propagator (b) from 2+1 flavor lattice QCD simulations, extended by functional computations in Yang-Mills theory and QCD and compared against the correlators obtained from the reconstructed spectral functions shown in \Cref{fig:spectral-functions}. The results agree within the given statistical uncertainties as shown in the bottom panels, where the posterior GPs for the correlators are evaluated at the fixed momenta provided by the lattice data, which is then subtracted leaving the error bars intact. The total mean squared errors amount to $\sim$5e--6 for the ghost and $\sim$4e--5 for the gluon.}
	\label{fig:correlators}
\end{figure*}

Conversely, the framework also enables inference in the opposite direction. The inherent analytic tractability associated with Gaussian statistics allows formulating the conditional distribution for $\rho(\omega)$ given observations $G_i$ in closed form. The full expression may then be derived as
\begin{equation}\label{eq:posterior}
\begin{aligned}
	\rho(\omega)\,|\,G_i &\sim \mathcal{GP} \Big(\mu(\omega)\ + \\
	\sum_{i,j=1}^{N_G} \int \textrm{d}\eta\, K(p_i,\eta) &C(\eta,\omega) \left(\tilde C + \sigma_n^2 \cdot \mathbf{1} \right)^{-1}_{ij} (G_j - \tilde\mu_j)\ , \\
	C(\omega,\omega') - \sum_{i,j=1}^{N_G} \int \textrm{d}&\eta\textrm{d}\eta'\, K(p_i,\eta) C(\eta,\omega) \\
	\left( \tilde C + \sigma_n^2 \cdot \mathbf{1} \right)^{-1}_{ij} &K(p_j,\eta') C(\eta',\omega') \Big) \ .
\end{aligned}
\end{equation}
The GP in \Cref{eq:posterior} encodes our knowledge of the spectral function after making observations of the propagator and accounting for observational noise with variance $\sigma_n^2$. The corresponding expressions for the dressing function instead of the propagator can be immediately obtained by inserting an additional factor of $p_i^2$ at every occurrence of the KL kernel $K(p_i,\omega)$ in \Cref{eq:prop-dist,eq:posterior}.

The flexibility of the approach makes it possible to also incorporate further available prior information in various forms into the predictive distribution in the same manner, yielding similar though somewhat more complicated expressions. This may include e.g.\ direct observations of $\rho$ and its derivatives, assumptions about the asymptotic behavior, or global normalization constraints.

In order for GPs to be useful for modeling, the covariance $C(\omega,\omega')$ may be defined via a so-called kernel function. It is commonly parametrized using a small number of hyperparameters, which may be subjected to optimization based on the associated likelihood. The mean function $\mu(\omega)$ is often set to zero, since its contribution can be fully absorbed by the kernel. Typically, the latter is the sole focus of the optimization procedure. However, a custom mean function may still be useful in certain situations in order to incorporate prior beliefs about the functional form of the expected solution, which can improve the calculation by providing a better starting point for the optimization routine.

A frequently used kernel parametrization is the radial basis function (RBF) kernel, also called squared exponential. It is defined as
\begin{equation}
    C(\omega,\omega') = \sigma_C^2\, \exp(-\frac{(\omega - \omega')^2}{2l^2})\ ,
\end{equation}
where the parameter $\sigma_C$ controls the overall magnitude and $l$ is a generic length scale. The RBF kernel has been established as the standard choice for many applications due to a number of attractive features, such as universality \cite{JMLR:v7:micchelli06a} and every function in its prior being infinitely differentiable. It is also used for our first results on spectral reconstruction with GPR presented in this work.

\begin{figure*}
	\centering
	\subfloat[]{%
		\includegraphics{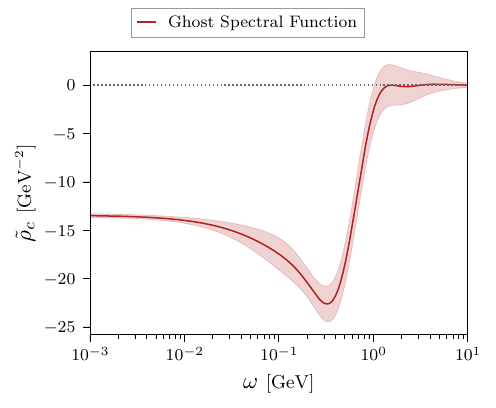}\label{fig:ghost-spectral}}%
	\hfill
	\subfloat[]{%
		\includegraphics{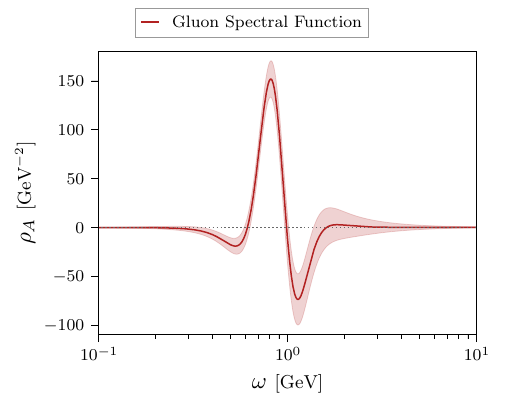}\label{fig:gluon-spectral}}%
	\caption{Plots showing the continuous part of the ghost (a) and the gluon spectral function (b) reconstructed from the lattice QCD correlators shown in \Cref{fig:correlators} using GPR. Shaded areas represent the $1\sigma$-bands of plausible solutions around the mean prediction based on the available observations and precision. The ghost spectral function $\rho_c$ features an additional massless particle pole in the origin; cf.~\Cref{eq:rho-ghost}.}
	\label{fig:spectral-functions}
\end{figure*}

Nevertheless, designing custom kernels for specific problems has been shown to greatly increase the usefulness of the approach in various settings and is also promising here. In particular, it may be interesting to construct kernel functions that can be integrated analytically against the KL kernel, such that the frequency integrals in \Cref{eq:prop-dist,eq:posterior} may be carried out analytically instead of numerically. To this end, one could potentially employ functions of Breit-Wigner type as done for the spectral function itself in~\cite{Cyrol:2018xeq}. In contradistinction, we may use them to instead define a suitable GP kernel, thereby still avoiding the restriction to a specific functional basis as previously mentioned. We comment on this and other possible improvements to our reconstruction approach in the conclusion.

Furthermore, we emphasize that the present approach in principle does not require us to choose a specific set of nodes $\omega_i$. In fact, instead of computing a discrete set of point predictions or coefficients of a predefined functional basis, the prediction for $\rho$ is obtained as a function of $\omega$, albeit only implicitly via the kernel formulation. In particular, the GP also allows computing all of the derivatives of the prediction analytically at any point---including the associated statistical uncertainties---by differentiating the expressions in \Cref{eq:posterior} with respect to $\omega$ (as well as $\omega'$ for the covariance). A finite set of nodes $\omega_i$ is chosen only at inference time in order to evaluate the GP, however, the choice is completely arbitrary within the given domain. This property is one of the most attractive features of GPR for spectral reconstruction and probabilistic function prediction in general.\\[-1ex]


\paragraph*{Input Data.}
In the past two decades, increasing interest in the momentum behavior of the fundamental two-point Green's functions in QCD as well as further correlation functions of higher order has triggered respective lattice calculations in particular of Yang-Mills and QCD propagators; see e.g.~\cite{Bonnet:2000kw, Sternbeck:2005tk, Boucaud:2005gg, Silva:2005hb, Cucchieri:2006tf, Cucchieri:2008qm, Oliveira:2008uf, Bogolubsky:2009dc, Iritani:2009mp, Ayala:2012pb, Athenodorou:2018rsk, Duarte:2016ieu, Aguilar:2019uob, Aguilar:2021lke, Aguilar:2021okw}. The lattice data for the ghost dressing function and gluon propagator employed in this work are shown in \Cref{fig:correlators}. They are obtained from recent simulations of 2+1 flavor QCD at the physical point~\cite{Zafeiropoulos:2019flq, Cui:2019dwv}; see \Cref{app:data-lattice} for further details and references. Additional input data and benchmarks are provided by one-parameter families of solutions from functional computations in Yang-Mills theory and QCD~\cite{Cyrol:2016tym, Cyrol:2018xeq, Gao:2021wun, Horak:2021pfr}, which are matched to the continuum-extrapolated lattice data as shown in \Cref{fig:dressing-comparison,fig:spectral-comparison}; see \Cref{app:data-fun} for details.


\paragraph*{Reconstruction Results.}
The GPR for the reconstruction of the ghost spectral function is performed using the aforementioned standard RBF kernel. We extend the lattice input data for the dressing function into the deep IR and simultaneously fix the low-frequency asymptotics of the spectral function using a direct real-time result in Yang-Mills theory obtained via the spectral ghost DSE~\cite{Horak:2021pfr} (see also \Cref{app:data-fun}). This is achieved by treating the spectral DSE result as an additional observation. Our procedure uniquely determines the non-zero value of $\rho_c$ for $\omega \to 0^+$, but also increases the reliability of the solution in the most interesting central region with respect to the kernel hyperparameters. Using just the lattice data without the extension by the spectral DSE result leads to a much higher variance in the solution space, with widely different asymptotic behaviors of solution candidates in the IR. The kernel hyperparameters are chosen by optimizing the associated likelihood of observations with an additional Gaussian hyperprior, which we achieve through a fine-grained grid scan; see \Cref{app:implementation} for details. The reconstructed spectral function in \Cref{fig:ghost-spectral} accurately reproduces the dressing function data within the uncertainties displayed in \Cref{fig:ghost-dressing}, with a total mean squared error of $\sim$5e--6.

The features of our prediction are strikingly similar to the aforementioned Yang-Mills result shown in \Cref{fig:ghost-spectral-comparison} in \Cref{app:data}, even though only the IR limit is incorporated into the reconstruction. This is expected heuristically, since the ghost only interacts with the quarks indirectly via the gluon vertices, and the effects of introducing dynamical quarks must hence be of higher order. The similarity is particularly notable considering that the methods are conceptually very different.

For the reconstruction of the gluon spectral function, the lattice input data are extended into the UV using an earlier fRG computation~\cite{Cyrol:2018xeq}, which is quantitatively reliable in this regime. We discuss this in more detail in the next paragraph and in \Cref{app:data-fun}. As for the ghost, this extension leads to greatly enhanced stability of the reconstruction with respect to the kernel hyperparameters. In particular, it ensures convergence to zero for $\omega \to \infty$, whereas with just the lattice data we often observe convergence to a non-zero constant and in some cases even pathological divergences. A modified frequency scale is used in the RBF kernel in order to suppress spurious oscillations in the IR and UV tails.  The hyperparameters are again obtained via optimization of the likelihood with Gaussian hyperpriors while approximately enforcing the OZS condition; see \Cref{app:implementation} for details. The reconstruction shown in \Cref{fig:gluon-spectral} accurately reproduces the lattice data within the given uncertainties, as shown in \Cref{fig:gluon-propagator}, with a total mean squared error of $\sim$4e--5. While also being fully consistent, deviations from the lattice propagator are somewhat stronger than for the ghost dressing function and seem to become more pronounced in the IR. This is likely caused by the comparably large uncertainties of the lattice data at small momenta.

The peak structure of the spectral function appears similar to an earlier reconstruction of the Yang-Mills propagator in the fRG framework~\cite{Cyrol:2018xeq}, shown in \Cref{fig:gluon-spectral-comparison} in \Cref{app:data}. We emphasize that the UV extension is done with the Yang-Mills data of~\cite{Cyrol:2016tym} instead of the full 2+1 flavor results from~\cite{Gao:2021wun}. This is detailed in \Cref{app:data-fun} and facilitates the comparison with the Yang-Mills reconstruction~\cite{Cyrol:2018xeq}. In particular, the positions of the leading positive peaks approximately coincide, with $\omega \approx 0.818$ for the present result and $\omega \approx 0.835$ for the fRG reconstruction. This reflects the approximate coincidence of the peaks of the Euclidean gluon dressing functions shown in \Cref{fig:ghost-dressing-comparison} in \Cref{app:data}. We also note that a small peak to the right of the second local minimum is present in both reconstructions. This feature may be a generic reconstruction artifact since it is not necessitated by theoretical considerations, but is observed in both results from conceptually very distinct methods. However, the comparably large uncertainties in this region also include plausible solutions without additional zero-crossings.

Significant differences between the two reconstructions are observed mainly in the overall peak height and width. Generally, the QCD result for the gluon is expected to differ more strongly from the pure gauge theory than the ghost due to the direct coupling to quarks. However, differences may also be attributed in part to the limited availability and precision of data and the resulting difficulty in resolving highly peaked structures. We find that generating narrower peaks with greater amplitudes by allowing the kernel's magnitude parameter $\sigma_C$ to increase and the length scale $l$ to decrease leads to stronger oscillations in the solution. This is a common feature of conceptually similar reconstruction approaches, such as linear regression with a Tikhonov regularizer (also called ridge regression), which has been applied e.g.\ in~\cite{Dudal:2019gvn}. Introducing such a regularization scheme, which is equivalent to assuming a Gaussian prior, leads to a favoring of solutions that are closer to zero. This additional bias can introduce the unwanted oscillations. Within the GPR approach, the kernel hyperparameters provide more detailed control over the regularization and can be tuned to deliberately suppress such unphysical features. However, this may result in reconstructions that are naturally flatter, which must be taken into account when interpreting and utilizing the result. This demonstrates one of the key advantages of GPR, namely the possibility to dynamically adjust the resolution depending on the available amount and quality of the input data, while still matching the observations as accurately as possible.

Although the obtained spectral functions reproduce the lattice data to high accuracy, the asymptotic behaviors of the mean predictions in the deep IR and UV differ from the analytic results derived in \cite{Cyrol:2018xeq}. In particular, different scaling exponents are observed and the gluon spectral function shows the opposite sign in the UV. Nevertheless, the analytically expected behavior is still plausibly contained within the computed errors, which are comparably large in these regimes. This indicates that not enough prior information is available to the GP from just the data in order to accurately resolve the tails of the spectral functions, which may come as no surprise. While this issue does not affect the reconstruction in the region of interest, it may be problematic for precision computations that use these results as inputs. In order to directly enforce the correct asymptotics, potential approaches are the incorporation of the analytically known behaviors into the prior means of the GPs or finding more suitable choices for the kernel functions. Furthermore, exploiting the available analytic results to provide additional prior information about the derivative structure may be particularly helpful in stabilizing the tail behavior. To achieve this, one may again write down the joint distribution of the predicted spectral function at any frequency and its associated derivatives to arbitrary order in closed form and derive the conditional posterior distribution similar to \Cref{eq:posterior}.\\[-1ex]


\paragraph*{Conclusion.}
In this work, we apply Gaussian process regression to the reconstruction of ghost and gluon spectral functions in 2+1 flavor QCD at the physical point. These spectral functions are the pivotal building blocks of diagrammatic representations for bound state equations such as Bethe-Salpeter and Faddeev equations, see e.g.~\cite{Cloet:2013jya, Eichmann:2016yit, Sanchis-Alepuz:2017jjd}, as well as transport coefficients, see e.g.~\cite{Haas:2013hpa, Christiansen:2014ypa}. 

Importantly, the gluon spectral function has a pronounced quasi-particle peak, the position of which is related to the mass gap in QCD. This extends previous vacuum and finite-temperature results in Yang-Mills theory~\cite{Haas:2013hpa, Cyrol:2018xeq} to physical QCD. Our findings provide non-trivial QCD support to the phenomenological use of quasi-particle gluon spectral functions for transport computations; see~\cite{Bluhm:2020mpc} for a recent review. Moreover, the present results can be directly employed as first-principle QCD inputs in order to systematically improve the respective phenomenological approaches towards a first-principle treatment of QCD transport processes.

These promising phenomenological applications of the present results also highlight the necessity of further improving the reconstruction approach itself, for which a number of potential directions can be envisaged. This includes the aforementioned possibility of designing custom kernels for the problem at hand, potentially with analytic integrability against the KL kernel. Constructing suitable, expressive kernels may also be automated and improved through the use of hyperkernels~\cite{JMLR:v6:ong05a} or techniques such as deep kernel learning~\cite{wilson2015deep}. To account for some variability in the kernel hyperparameters, one may replace the maximum likelihood approach by an integral over parameter space using a suitable hyperprior which encodes any prior assumptions. Alternatively, optimal hyperparameters may also be selected based on a data-driven machine learning approach, using datasets consisting of pairs of correlators and associated spectral functions.

Furthermore, the flexibility of the GPR framework allows the incorporation of various supplementary constraints derived from theoretical arguments, such as information about derivatives, known asymptotic behaviors, or normalization conditions. This is expected to further improve the accuracy and reliability of the reconstruction, in particular for the IR and UV tails of the spectral functions that are otherwise difficult to resolve. This will be the subject of future work, accompanied by direct functional computations of further spectral properties along the lines of~\cite{Horak:2020eng, Horak:2021pfr}.

The immediate next steps in our endeavor towards unveiling real-time properties of QCD are the application and extension of the present numerical method to quark propagators as well as correlation functions computed at finite temperature. This will enable quantitative studies of hitherto theoretically inaccessible non-equilibrium dynamics of QCD in the transport phase of heavy-ion collisions within a first-principle approach.\\


\paragraph*{Acknowledgements.}
We thank A.~Cyrol, F.~Gao, L.~Kades, J.Y.~Lin, J.~Papavassiliou and A.~Rothkopf for discussions. We thank P.~Boucaud and F.~De~Soto for an earlier collaboration and for their help in the preparation of the lattice data. We are indebted to the RBC/UKQCD collaboration, especially to P.~Boyle, N.~Christ, Z.~Dong, N.~Garron, C.~Jung, B.~Mawhinney and O.~Witzel, for access to the lattices used in this work. We also thank the members of the fQCD collaboration \cite{fQCD} for discussions and collaborations on related subjects. JRQ acknowledges the support of MICINN PID2019-107844GB-C22 grant. Numerical computations have used resources of CINES, GENCI IDRIS (project id 52271) and of the IN2P3 computing facility in France. This work is supported by the Deutsche Forschungsgemeinschaft (DFG, German Research Foundation) under Germany's Excellence Strategy EXC 2181/1 - 390900948 (the Heidelberg STRUCTURES Excellence Cluster), under the Collaborative Research Centre SFB 1225 (ISOQUANT), EMMI, and the BMBF grant 05P18VHFCA.

\appendix


\section{Introduction to GPR}
\label{app:gpr}

This appendix serves as a brief introduction to GPR for function prediction using a finite number of direct or indirect observations, based primarily on~\cite{10.1093/gji/ggz520}. We adopt the notation used in the main text for consistency, however, the general formalism presented here is also applicable outside of the specific context of spectral reconstruction for quantum field theory. For a modern, comprehensive textbook treatment of the topic, we refer the interested reader to~\cite{10.5555/1162254}. For a brief, pedagogical introduction to GPR with simple code examples, we recommend~\cite{krasser_2018}. In the context of inverse theory, \cite{Menke2021} provides a recent review.

We first discuss GPR for the case where direct observations are available for the function to be modeled. We assume our knowledge of the function $\rho(\omega)$ to be encoded in a GP with mean and covariance functions $\mu(\omega),C(\omega,\omega')$, denoted by
\begin{equation}
    \rho(\omega) \sim \mathcal{GP}\left(\mu(\omega),C(\omega,\omega')\right) \ ,
\end{equation}
where the covariance is assumed to be symmetric, i.e.\ $C(\omega,\omega') = C(\omega',\omega)$. As per the definition of a GP, any finite set of function evaluations at $N$ sample points $\omega_i$ follows a multivariate normal distribution,
\begin{equation}
\begin{aligned}
  \begin{pmatrix}
    \rho(\omega_1) \\
    \vdots \\
    \rho(\omega_N)
  \end{pmatrix}
  & \sim \mathcal{N} \left(
  \begin{pmatrix}
    \mu(\omega_1) \\
    \vdots \\
    \mu(\omega_N)
  \end{pmatrix}, \right. \\
  & \left. \begin{pmatrix}
    C(\omega_1,\omega_1) & \dots & C(\omega_1,\omega_N) \\
    \vdots & \ddots & \vdots \\
    C(\omega_N,\omega_1) & \dots & C(\omega_N,\omega_N)
  \end{pmatrix} \right) \ .
\end{aligned}
\end{equation}

Similarly, we can write down the joint distribution of a set of observations $\hat{\rho}_i$ at points $\hat{\omega}_i$ and the value of the function at an arbitrary point $\omega$ as
\begin{equation}
    \begin{pmatrix}
      \rho(\omega) \vspace{1ex} \\
      \bm{\hat{\rho}}
    \end{pmatrix} \sim \mathcal{N} \left(
    \begin{pmatrix}
      \mu(\omega) \vspace{1ex} \\
      \bm{\hat\mu}
    \end{pmatrix},
    \begin{pmatrix}
      C(\omega,\omega') & \mathbf{\hat{C}}^T(\omega) \vspace{1ex} \\
      \mathbf{\hat{C}}(\omega') & \mathbf{\hat{C}} + \sigma_n^2 \cdot \mathbf{1}
    \end{pmatrix}\right) \ ,
\end{equation}
where boldface type denotes vector and matrix quantities. Here, we have defined $\bm{\hat{\mu}} \equiv \mu(\hat{\omega}_i)$, $\mathbf{\hat{C}}_i(\omega) \equiv C(\hat{\omega}_i,\omega)$, and $\mathbf{\hat{C}}_{ij} \equiv C(\hat{\omega}_i,\hat{\omega}_j)$. $\sigma_n^2$ defines the point-wise variance of additional measurement noise which may be present in the observations $\bm{\hat{\rho}}$. Due to the analytic tractability of multivariate Gaussians, the conditional distribution of function values $\rho(\omega)$ given observations $\bm{\hat{\rho}}$ may then be derived as
\begin{equation}
\begin{aligned}
    \rho(\omega) | \bm{\hat{\rho}} \sim \mathcal{N}\left(\mu(\omega) + \mathbf{\hat{C}}^T(\omega) \left(\mathbf{\hat{C}} + \sigma_n^2 \cdot \mathbf{1}\right)^{-1} (\bm{\hat{\rho}} - \bm{\hat{\mu}}) \right.,  \\
    \left. C(\omega,\omega') - \mathbf{\hat{C}}^T(\omega) \left(\mathbf{\hat{C}} + \sigma_n^2 \cdot \mathbf{1}\right)^{-1} \mathbf{\hat{C}}(\omega') \right) \ .
\end{aligned}
\end{equation}
\\[-1ex]
The covariance is parametrized by a suitable kernel function, whereby one may encode any prior beliefs about the types of solutions one expects by choosing an appropriate form for the problem at hand. For an introduction to constructing GP kernels of various types as well as strategies to apply and combine them, we recommend the kernel cookbook~\cite{duvenaud}.

A kernel's hyperparameters, denoted here by $\bm{\hat{\alpha}}$, may be subjected to optimization by maximizing the associated likelihood,
\begin{equation}
\begin{aligned}
    p&(\bm{\hat{\rho}} | \bm{\alpha}) = \left( (2\pi)^N \det\left(\mathbf{\hat{C}}_\alpha + \sigma_n^2 \cdot \mathbf{1}\right) \right)^{-\frac{1}{2}} \cdot \\
    &\exp\left( -\frac{1}{2} (\bm{\hat{\rho}} - \bm{\hat{\mu}})^T \left( \mathbf{\hat{C}}_{\bm{\alpha}} + \sigma_n^2 \cdot \mathbf{1} \right)^{-1} (\bm{\hat{\rho}} - \bm{\hat{\mu}}) \right)\ ,
\end{aligned}
\end{equation}
where we have written $\mathbf{\hat{C}}_{\bm{\hat{\alpha}}}$ to emphasize the dependence on the hyperparameters. Instead of directly maximizing $p(\bm{\hat{\rho}} | \bm{\alpha})$ as a function of $\bm{\hat{\alpha}}$, one conventionally minimizes the negative log likelihood (NLL),
\begin{equation}\label{eq:nll}
\begin{aligned}
    -\log &\, p(\mathbf{\hat{f}} | \bm{\alpha}) = \frac{1}{2} (\bm{\hat{\rho}} - \bm{\hat{\mu}})^T \left( \mathbf{\hat{C}}_{\bm{\alpha}} + \sigma_n^2 \cdot \mathbf{1} \right)^{-1} (\bm{\hat{\rho}} - \bm{\hat{\mu}}) \\
    & + \frac{1}{2} \log \det \left(\mathbf{\hat{C}}_\alpha + \sigma_n^2 \cdot \mathbf{1}\right) + \frac{N}{2} \log 2\pi\ .
\end{aligned}
\end{equation}
Since simply finding and employing the maximum likelihood configuration of hyperparameters may ignore relevant additional structures in the distribution, one can also integrate out $\bm{\hat{\alpha}}$ using suitable hyperpriors to account for some variability.

Based on the formulation of GPR for direct observations $\bm{\hat{\rho}}$ at points $\bm{\hat{\omega}}$, one can derive the expressions for inference from indirect observations $\mathbf{\hat{G}}$ at points $\mathbf{\hat{p}}$ as discussed in the main text by applying the forward process of the associated linear inverse problem, in our case the KL integral defined in \Cref{eq:KL_rep}. This involves all terms related to the observations that depend on the discrete set of points $\bm{\hat{\omega}}$, which are promoted back to the continuous domain and subsequently integrated out to yield the nodes $\mathbf{\hat{p}}$ instead.


\section{Input Data}
\label{app:data}

Combining the data from lattice simulations and functional computations as described in the main text requires matching the scales through renormalization. In this work, we always rescale the functional methods results to match the lattice data in the appropriate regime.

\begin{figure*}
    \centering
    \subfloat[]{%
     \includegraphics{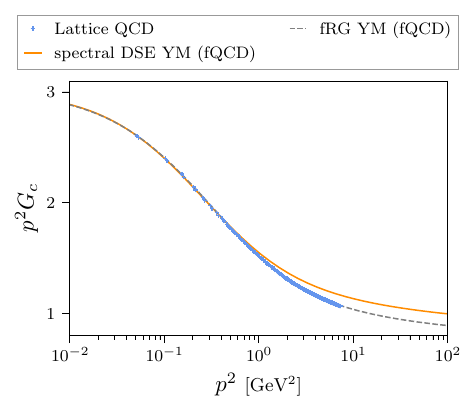}\label{fig:ghost-dressing-comparison}}
    \hfill
    \subfloat[]{%
    \includegraphics{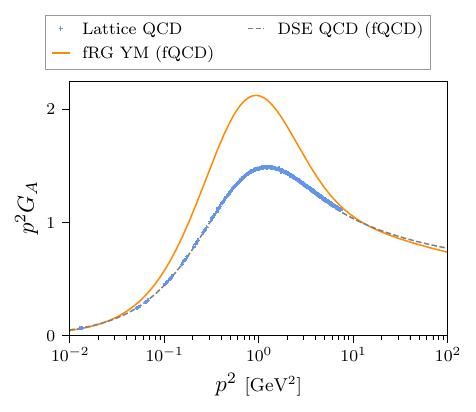}\label{fig:gluon-dressing-comparison}}%
    \caption{Plots showing ghost (a) and gluon (b) dressing functions in 2+1 flavor QCD and Yang-Mills (YM) theory, obtained from the lattice simulations and functional computations discussed in \Cref{app:data}.}
    \label{fig:dressing-comparison}
\end{figure*}

\begin{figure*}
    \centering
    \subfloat[]{%
    \includegraphics{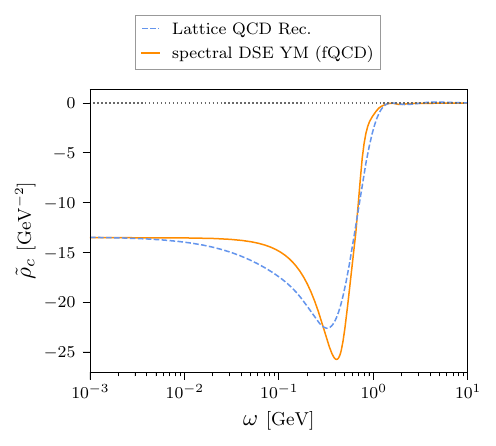}\label{fig:ghost-spectral-comparison}}%
    \hfill
    \subfloat[]{%
    \includegraphics{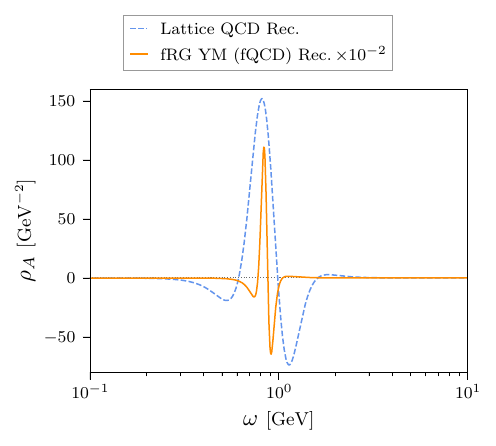}\label{fig:gluon-spectral-comparison}}%
    \caption{Plots comparing the continuous part of the ghost (a) and the gluon spectral function (b) from different approaches in 2+1 flavor QCD and Yang-Mills (YM) theory, as discussed in the results section and \Cref{app:data}. The ghost spectral function $\rho_c$ features an additional massless particle pole in the origin; cf.~\Cref{eq:rho-ghost}.}
    \label{fig:spectral-comparison}
\end{figure*}


\subsection{Lattice Simulations}
\label{app:data-lattice}

The lattice data employed in this work were obtained from configurations generated by the RBC/UKQCD collaboration---first introduced in~\cite{Allton:2007hx, Allton:2008pn, Arthur:2012yc, Blum:2014tka, Boyle:2017jwu}---with 2+1 dynamical quark flavors using the Iwasaki~\cite{Iwasaki:1985we} and domain wall fermion~\cite{Kaplan:1992bt, Shamir:1993zy} actions, respectively for the gauge and quark sectors, at the physical point (a pion mass amounting to 139 MeV) by the particular implementation of the M\"obius kernel~\cite{Brower:2004xi}.  These developments were then exploited in~\cite{Zafeiropoulos:2019flq, Cui:2019dwv} in order to calculate the gluon and ghost propagators as well as the strong coupling in a particular scheme~\cite{Sternbeck:2007br, Boucaud:2008gn, Sternbeck:2010xu}, and an effective charge stemming from it~\cite{Binosi:2016nme}. A description of this calculation is given, for instance, in~\cite{Ayala:2012pb}.

In computing propagators that properly feature the physical running with momenta, data should be thoroughly cured from lattice regularization artifacts. In particular, as explained in~\cite{Zafeiropoulos:2019flq}, our results are obtained after a careful scrutiny of discretization artifacts, thereby accounting for the continuum-limit extrapolation, following~\cite{Boucaud:2018xup}. As a noteworthy remark, a recent work~\cite{Aguilar:2021okw} has revealed the key role played by the procedure of~\cite{Boucaud:2018xup} for an adequate removal of discretization artifacts in achieving a consistent description of Yang-Mills two- and three-point correlators, involving both lattice and DSE results.

The resulting ghost dressing function and gluon propagator data are displayed in \Cref{fig:ghost-dressing,fig:gluon-propagator}, respectively. They are compared against their counterparts obtained from evaluating \Cref{eq:KL_rep} for the reconstructed spectral functions shown in \Cref{fig:spectral-functions}, as well as the results from functional methods described in the following section. The dressing functions of all input datasets are compared in \Cref{fig:dressing-comparison} to further illustrate their similarities and differences.


\subsection{Functional Methods}
\label{app:data-fun}

We briefly summarize results from functional computations in Yang-Mills theory and QCD that are employed in this work to provide additional prior information for the reconstruction. For reviews on the application of functional methods in this context, see e.g.~\cite{Binosi:2009qm, Huber:2018ned, Fischer:2018sdj, Dupuis:2020fhh}. 

We use the real-time Yang-Mills results from~\cite{Horak:2021pfr} to extend the lattice QCD data of the ghost dressing function into the deep IR, as shown in \Cref{fig:ghost-dressing-comparison}. The approach also provides direct access to the associated spectral function, which we employ to fix the low-frequency asymptotic behavior of the reconstruction. It is obtained via the spectral ghost DSE, building upon the technique of spectral renormalization~\cite{Horak:2020eng}. Making use of \Cref{eq:KL_rep} for the ghost and gluon propagator, the momentum integrals appearing in the loop diagrams of the ghost propagator DSE can be solved analytically. This preserves the full analytic momentum dependence and allows evaluating the equation on the real momentum axis. The spectral function can then be directly extracted from the real-time propagator DSE via \Cref{eq:rho-def}; see \Cref{fig:ghost-spectral-comparison} for a comparison to the reconstruction result of the present work. As input gluon spectral function, the reconstruction result of~\cite{Cyrol:2018xeq} based on the scaling solution obtained via the fRG in~\cite{Cyrol:2016tym} is used. Assuming a spectral representation for the gluon propagator, in both scaling and decoupling scenario the IR behavior of the gluon spectral function follows directly from the propagator~\cite{Cyrol:2018xeq}. This is utilized to modify the given scaling spectral function such that we obtain a decoupling-type gluon propagator matching the value of the given lattice propagator well within the given uncertainties. Due to its mild momentum dependence, the ghost-gluon vertex is assumed to be classical.

The lattice QCD data for the gluon propagator are extended towards the UV using earlier results from functional computations in Yang-Mills theory~\cite{Cyrol:2016tym}. Differences to the 2+1 flavor QCD result for the gluon propagator reported in~\cite{Gao:2021wun}, being based on~\cite{Cyrol:2017ewj}, are comparably small in the relevant momentum range. A stronger deviation can be observed in the dressing functions, as shown in \Cref{fig:gluon-dressing-comparison}. Despite these differences, the reconstruction still produces remarkably reliable results, cf.~\Cref{fig:gluon-propagator}. Nevertheless, we aim to replace the Yang-Mills UV extension by the 2+1 flavor QCD data from~\cite{Gao:2021wun} in order to further optimize the accuracy of the result and mitigate any potential issues. For related results and further correlation functions see~\cite{Aguilar:2012rz, Williams:2015cvx, Fu:2019hdw, Gao:2020qsj}. More specifically, the fRG results in~\cite{Cyrol:2016tym} are derived within an advanced approximation where the momentum dependence of all vertices is approximated at the symmetric point, for respective DSE results see~\cite{Huber:2020keu}. For our purposes, this data set provides the optimal trade-off for momentum range versus accuracy. Due to the high numerical precision, the results are particularly well-suited as an input for spectral reconstruction. The Yang-Mills data have already been employed for this purpose in~\cite{Cyrol:2018xeq} and we use this earlier reconstruction for comparison; see \Cref{fig:gluon-spectral-comparison}. In summary, the extension of the 2+1 flavor lattice data with the high precision Yang-Mills data up to momenta $p^2 = 10^2$\,GeV${}^2$ allows a more direct comparison (in terms of scales) with the Yang-Mills reconstruction in~\cite{Cyrol:2018xeq}, while only modifying the large frequency tail of the gluon spectral function for frequencies $\omega \gtrsim 5$\,GeV, see \Cref{fig:spectral-comparison}.


\section{Implementation}
\label{app:implementation}

In this section, we comment on certain points of the implementation in more detail. We first address numerical aspects of the optimization and a discussion of the required computational effort. Subsequently, we provide further information about data usage, kernel design choices and theoretical constraints for the particular reconstructions reported in this work.


\subsection{Hyperparameter Optimization and Computational Cost}

To find optimal values for the kernel's hyperparameters, we perform a fine-grained grid scan of the NLL with additional hyperpriors where necessary. Alternatively, the NLL may also be minimized with a gradient-based ansatz using a standard optimizer such as L-BFGS. However, mapping out the posterior distribution in more detail tends to be highly instructive for the problem at hand. It is also less prone to numerical problems such as unstable directions and violation of positive definiteness of the covariance, as these can be identified early on, and should hence be preferred when feasible. This is also where the bulk of the computational effort goes, as it involves calculating for each individual grid point the comparably expensive inverse and determinant of the covariance matrix, which naively scales like $\mathcal{O}(N^3)$. For very large datasets where their direct evaluation becomes infeasible, one may resort to cheaper linear solvers for the inverse and stochastic approximations of the determinant, but this is unlikely to become necessary in this particular context. Cost may also be mitigated by scanning the parameter space hierarchically, starting at low resolution and zooming into the interesting regions.

The whole procedure is trivially parallelizable, as each grid point can be treated independently. At the scale of the present work, each instance was handled by a standard CPU node with low performance requirements. Some first tests were also conducted on a single machine, where mapping out the parameter space for each reconstruction with medium resolution took a few hours at most. In comparison to finding the optimal hyperparameters, the subsequent inference step is negligibly cheap. Of course, the total computational effort for the reconstruction is dwarfed by the requirements of the large-scale lattice simulations described in \Cref{app:data-lattice}, which are orders of magnitude more expensive.


\subsection{Reconstruction Details}

\begin{figure}
    \centering
    \includegraphics{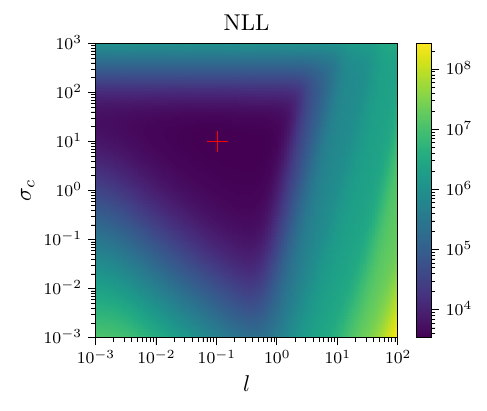}
    \caption{Heatmap of the NLL as a function of the RBF kernel hyperparameters $\sigma_C,l$ for the reconstruction of the ghost spectral function, with an additional zero-mean Gaussian hyperprior for $\sigma_C$. A unique minimum can be identified, which provides the optimal values used for the results shown in \Cref{fig:ghost-dressing,fig:ghost-spectral}.}
    \label{fig:ghost-nll}
\end{figure}

\subsubsection{Ghost}

In the case of the ghost spectral function, we treat the low-frequency asymptotics extracted from the direct DSE computation in Yang-Mills theory as an additional observation for the GP. This is only possible for the ghost, as a similarly direct determination of the Yang-Mills gluon spectral function is currently not available. The procedure is implemented by including the value of $\rho$ at $\omega = 0$ in the construction of the joint distribution of observations and predictions. In particular, one needs to compute additional expressions for the covariances of the point $\rho(0)$ and the correlator data. This requires some programming headache, but carries no further conceptual difficulty.

As stated in the main text, we use the standard RBF kernel and identify optimal hyperparameters via a high-resolution grid scan. We note an unstable direction in the magnitude parameter $\sigma_C$, which is cured by subjecting it to a zero-mean Gaussian hyperprior. As an illustrative example, the heatmap for the NLL including this additional regularization term for $\sigma_C$ is shown in \Cref{fig:ghost-nll}.

\subsubsection{Gluon}

In the case of the gluon spectral function, no real-time result in Yang-Mills theory is available to fix the asymptotics. However, as an additional theoretical constraint we require the solution to respect the aforementioned OZS condition defined in \Cref{eq:gen-ozs}. While one might expect this to further complicate the reconstruction, it actually helps in narrowing down the space of plausible solutions. The condition can simply be enforced approximately by treating it as an additional indirect observation and checking it a posteriori. The associated transformation is here just the convolution with $\omega$ instead of the KL integral. We confirm that the OZS condition is fulfilled with a relative accuracy of $\sim$1\%, computed by evaluating the ratio of the left-hand side of \Cref{eq:gen-ozs} and the same expression using the modulus of the integrand, i.e.\ $\int_0^\infty \textrm{d}\omega \, |\omega \rho_A(\omega)|$.

As mentioned in the main text, we find it helpful to modify the standard RBF kernel by non-linearly rescaling the frequency as $\omega \rightarrow \tilde{\omega} = \omega^4 (1 + \omega^4)^{-1}$ before computing the squared distance. This leads to a strongly improved asymptotic stability of the reconstructed spectral function, in particular at large frequencies, compared to just using $\omega$ itself. The procedure may be interpreted either as a non-stationary modification of the kernel or as a preprocessing step for the data to the same effect.


\newpage
\bibliographystyle{utphys}
\bibliography{literature.bib}

\end{document}